\documentclass[twocolumn,showpacs,preprintnumbers,amssymb]{revtex4}
\usepackage{graphicx}
\usepackage{dcolumn}
\usepackage{bm}

\begin{document}
\draft          
\title{Quasi-Scarred Resonances in a Spiral-Shaped Microcavity}
\author{Soo-Young Lee}
\author{Sunghwan Rim}
\author{Jung-Wan Ryu}
\author{Tae-Yoon Kwon}
\author{Muhan Choi}
\author{Chil-Min Kim}
\affiliation{National Creative Research Initiative Center for Controlling Optical Chaos,\\
Pai-Chai University, Daejeon 302-735, Korea}

\begin{abstract}
We study resonance patterns of  a spiral-shaped dielectric microcavity with 
chaotic ray dynamics. 
Many resonance patterns of this microcavity, with refractive indices
$n=2$ and $3$, 
 exhibit strong localization of simple geometric shape, and we call them 
{\em quasi-scarred resonances} in the sense that there is, unlike the conventional scarring,
no underlying periodic orbits. 
It is shown that the formation of quasi-scarred pattern can be understood in terms of
ray dynamical probability distributions and wave properties like uncertainty and interference.
\end{abstract}


\maketitle

\narrowtext

 The scar phenomenon, since its advent in a chaotic billiard, has attracted much attention~\cite{He84},
 because it had not been anticipated from the prevailed random matrix theory~\cite{Bo84}. 
It is now known that the scarred 
eigenfunctions show not only strong enhancement along a unstable
 periodic orbit, but also
detail of the stable and unstable manifolds around the periodic orbit~\cite{Cr02}.
This scar effect therefore has been regarded as an important
feature of chaotic systems different from random systems. Another important
aspect of the scarring effect is its ubiquitous existence; it has been observed
 in various chaotic systems such as microwave cavity~\cite{Sr91},
 semiconductor quantum well~\cite{Fr95}, surface wave~\cite{Ku01},
 optical cavities~\cite{Lee02,Re02}, etc.

Recently there is a considerable interest in the light emission from dielectric
cavities with chaotic ray dynamics, since many intriguing light emission behaviors
take place and are known to be relevant to the underlying chaotic ray dynamics~\cite{Gm98}.
There are several reports of observation of scarred lasing modes in dielectric
microcavities of various boundary shapes~\cite{Lee02,Gm02,Re02}.
The scarred lasing modes generally show good directionality of light emission,
the directionality is an important characteristic required for applications
to photonic and optoelectric information processing~\cite{Ch96}.
The number of directional beams of the scarred emission from usual microcavities
would be more than two because of the discrete symmetry of the boundary geometry
and the possibility of interchanging incident and reflected rays on the underlying
periodic orbit.

In a remarkable experiment, Chern et al. have successfully observed unidirectional emission 
in spiral-shaped quantum-well microlasers~\cite{Ch03}. The unidirectional laser beam is
important to arrange easy optical communication between microlasers.
The spiral-shaped boundary, in which ray dynamics is chaotic, is given by
\begin{equation}
  r(\phi)=R (1+ \frac{\epsilon}{2\pi} \phi)
\end{equation}
in polar coordinates ($r$, $\phi$), where $R$ is the radius of the spiral at $\phi=0$
and $\epsilon$ is the deformation parameter.
Basically, the unidirectionality of the emission beam comes from the special
properties of the spiral-shaped boundary geometry which other common cavity designs
do not have. They are the absence of any symmetry and the existence of the notch.
As mentioned above, the absence of symmetry would be the necessary condition
for the unidirectional emission.  The notch makes the microcavity show
very strong chirality by transmitting or reflecting counterclockwise rotating rays.
The bouncing from the notch is inevitable for the rays and would be an essential
process for the unidirectional emission. 
Besides the unidirectionality, it is important and interesting to study
how the unique characteristics of the spiral-shaped microcavity appear on resonance patterns.

In this Letter, we investigate the resonance patterns in the spiral-shaped dielectric microcavity.
We find that a large number of resonances obtained are strongly
localized and that the localized patterns are not supported by any unstable
periodic orbit, so we call them {\it quasi-scarred resonances}.
The existence of quasi-scarred resonances implies that the scarring phenomenon in dielectric
microcavities has substantial differences from the conventional
scarring in billiard systems. The differences come from inherent
characteristics of dielectric cavities such as existence of the
critical incident angle for total internal reflection and energy loss
by refractive emission. We explain the formation of the quasi-scarred
resonances in terms of ray dynamical probability distributions and
wave properties like uncertainty and interference. For convenience, we
take $\epsilon=0.1$ and $R=1$ in this Letter.

In order to investigate the ray dynamical properties of the spiral-shaped dielectric microcavity,
we first consider a uniform ensemble of initial points over the whole
phase space $(s,~p)$, where
$s$ is the boundary arc length from the $\phi=0$ point (see Fig. 4) and its conjugate variable
 $p$ is given as
$p=\sin \theta$,  $\theta$ being the incident angle of ray. If the boundary is made by a perfect
mirror, the distribution of the points in the phase space at later times would remain 
uniform (in a random sense) and structureless. However, in the dielectric microcavity,
the distribution of the points is, some time later, not uniform but
rather structural because the individual ray can suffer energy loss by
refractive emission when bouncing from boundary. The amount of the
energy loss is determined by the transmission coefficient ${\cal
  T}(p)$~\cite{Ha95} which has a nonzero value in the range of  $-p_c < p <
p_c$, where $p_c$ is the critical line for total internal reflection
and is related to the refractive index $n$ as $p_c=\sin \theta_c=1/n$,
$\theta_c$ being the corresponding critical incident angle. This leaky
property of rays in the ensemble is described by the {\it survival
  probability distribution} $\tilde{P}(s,p,t)$, the probability with
which the ray with $(s,~p)$ can survive in the microcavity at a time $t$. 
With the survival probability distribution $\tilde{P}(s,p,t)$, the energy ${\cal E} (t)$ 
confined in the microcavity and the {\it escape time distribution}
$P_{es} (t)$ are expressed as
 ${\cal E} (t) = {\cal E}_0 \int ds\, dp \, \tilde{P}(s,p,t)$, ${\cal E}_0$ being 
the initial energy, and $ P_{es} (t) =  \int ds\, dp \,
\tilde{P}(s,p,t) {\cal T}(p)$, respectively. Since the confined energy
decreases by the ray transmission through cavity boundary, we can get a relation,
\begin{equation}
\frac{d {\cal E} (t)}{d t} = -{\cal E}_0 P_{es} (t).
\label{e-p}
\end{equation}

\begin{figure}
\vspace{-0.7cm}
\hspace{-0.5cm} \includegraphics[height=6.2cm, width=7cm]{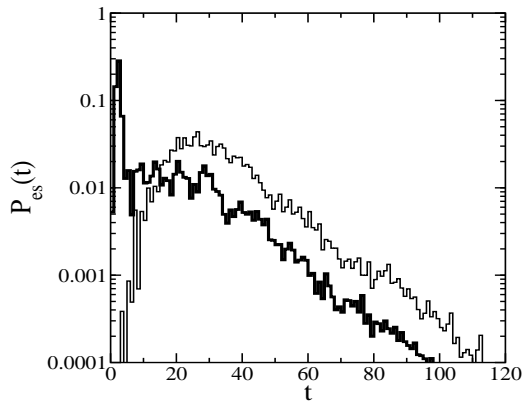}
\vspace{-0.5cm}
\caption{The escape time distributions $P_{es}(t)$ with $n=2$ for two different
sets of initial points; the thick and thin lines represent numerical results
 for Set A and Set B, respectively.
The time $t$ is scaled to be the length of ray trajectory when $R=1$.}
\vspace{-0.3cm}
\end{figure}

It is well known that in fully chaotic open systems the escape time distribution  
$P_{es} (t)$ shows exponential long time behavior, while it becomes power law decay in the KAM systems 
due to the stickiness of the KAM tori~\cite{Sc02}. The exponential behavior of  $P_{es} (t)$ suggests 
that $\tilde{P}(s,p,t)$ would have the same phase space distribution
after a certain period of time, i.e.,
\begin{equation}
\tilde{P}(s,p,t)=B(t)P_s(s,p),
\label{spd}
\end{equation}
which defines the {\it steady probability distribution} $P_s(s,p)$ as the stationary part 
of $\tilde{P}(s,p,t)$. It is obvious from Eq.~(\ref{e-p}) that the relation in Eq.~(\ref{spd})
is equivalent to assuming the exponential time behaviors of ray dynamical distributions such as
  ${\cal E} (t)$, $\tilde{P}(s,p,t)$, and $ P_{es} (t)$.
In the case of dielectric microcavities, a numerical justification of the relation in Eq.~(\ref{spd})
will be presented below (see Fig.~1).
The steady probability distribution $P_s(s,p)$ then characterizes the ray dynamical
long time behavior. The decay rate $\gamma$, from Eq.~(\ref{e-p}), can be expressed as
\begin{equation}
\gamma=\int ds\, dp \, P_s(s,p) {\cal T}(p),
\end{equation}
and the ray dynamical near field and far field distributions can be
also described by $P_s(s,p)$.

   For simplicity's sake, we will concentrate on TM (transverse magnetic) polarization 
in this Letter. 
 In Fig.~1, the escape time distributions $P_{es}(t)$ are shown for the $n=2$ case.
Here, we consider two different sets of $400 \times 400$ initial
points: one is the uniformly distributed set over the whole phase
space (Set A) and the other is the uniformly distributed one in a part
of the phase space, $(0< s <s_{max}/2,~0.5 < p < 0.75)$ (Set B),
 where $s_{max}$ is the total length of the boundary.
Note that above $t_c \simeq 30$ exponential decay behaviors are shown. 
The slope of the linear part determines the decay rate $\gamma$. 
The similar slopes for both Set A and Set B reflect that rays lose their energy  through
the same process.
The details of the process appear in the structure of $P_s(s,p)$.

\begin{figure}
\vspace{-0.2 cm}
\hspace{-0.6 cm} \includegraphics[height=4.5cm, width=8.8cm]{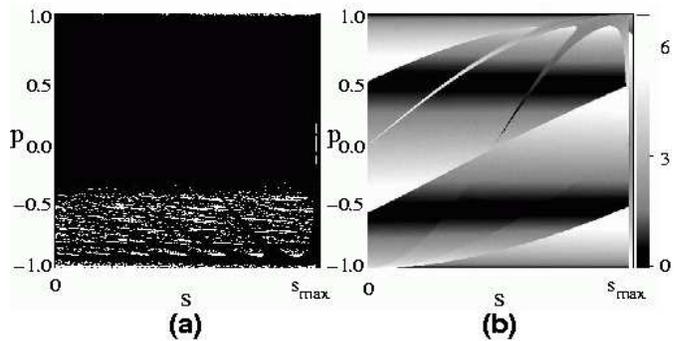}
\vspace{-0.2 cm}
\caption{(a) The normalized survival probability distribution $\tilde{P}(s,p,t)$
sampled in the time range of $57 < t <60$ for Set A.
The white dots represent points whose weight is greater than $0.1$ after normalization.
(b) The distance, resulting after 3 bounces, distribution $d(s,p)$.
From comparing both figures, it is clear that the partially reflected rays near $-p_c$
would make rough triangular trajectories. }
\vspace{-0.3cm}
\end{figure}

  Figure 2 (a) shows an approximate $P_s (s,p)$ for $n=2$ given by
normalizing the $\tilde{P}(s,p,t)$ in the time range of $57<t<60$
for Set A. The structure of the approximate $P_s(s,p)$ is almost invariant 
in other time ranges of the linear part $(t > t_c)$ and even for Set B.     
It is clear that the energy loss was mainly caused by tangential emissions
just above the critical line ($-p_c=-1/2$). 
So, we can see that the process mentioned above is the way that the ray trajectories first 
rotate counterclockwise ($p> p_c$), then change their rotational direction by reflection 
on the notch part, and afterwards gradually approach $-p_c$. Most of them 
are emitted out from the microcavity and the remains repeat the same process.
The distribution confined to the negative value of $p$ means strong chirality of
 this spiral-shaped microcavity. 
The dark tentacular structure in Fig.~2 (a) implies the missing trajectories
which are reflected at the notch with  $|p| > p_c$.
In fact, the overall structure presents a part of unstable manifolds, which is
typical in open chaotic systems~\cite{Sc02}.
This structure would give important informations about statistical properties of resonances, 
i.e., far field and near field distribution of resonances would show minima at
values corresponding to the missing trajectories.
 
  More direct implication on resonance patterns can arise from the distribution of 
resulting distance after 3 bounces (Fig.~2(b)), i.e., $d(s,p)=\sqrt{(s_f-s)^2 +(p_f-p)^2}$ where
$(s,~p)$ is the initial position and $(s_f,~p_f)$ being the position after 3 bounces.
We note that in Fig.~2 (b) the critical line $p=-p_c$ lies on the region of lower $d$ values. 
Since the rays in the region just above $-p_c$ are partially emitted out,  
the remaining reflected rays would make a rough triangle. As discussed below, the imprint of
this fact appears apparently in resonance patterns (see Fig.~3 (a)). 
Although the $n=3$ case is not presented in the figures, the distance distribution after
5 bounces also shows similar features, implying that the star shape ray trajectories
would be responsible for resonance patterns (see Fig.~3 (b)).

\begin{figure}
\begin{center}
\vspace{-0.2 cm}
\hspace{0. cm} \includegraphics[height=5.2cm, width=8.5cm]{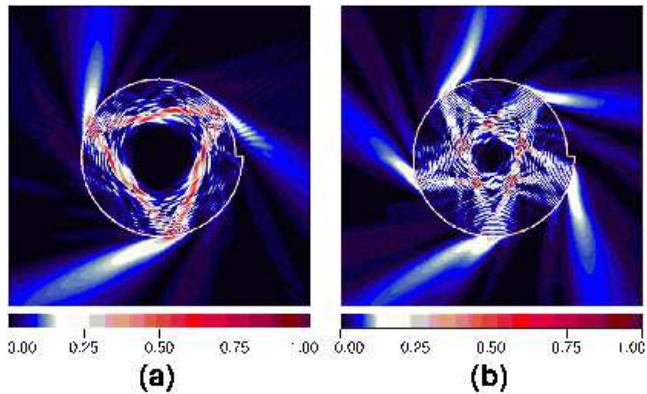}
\vspace{-0.5cm}
\caption{(color). Field intensity plots of quasi-scarred resonances in the 
spiral-shaped microcavity.
(a) $n=2$ and $nkR=(109.70,-0.1128)$. (b) $n=3$ and $nkR=(109.59,-0.1127)$.
In figures, the field intensity is normalized by scaling the maximum intensity as one.}
\end{center}
\vspace{-0.5cm}
\end{figure}

Using the boundary element method~\cite{Wie03}, we obtain resonances 
around $\mbox{Re}(nkR) \simeq 110$ for
the spiral-shaped dielectric microcavity, 24 resonances for $n=2$ and 23 resonances for $n=3$,
which are about 25\% of the total number of resonances in the concerned range. 
From the resonances we realize an important fact
that the basic localized structures of the resonance patterns are 
triangular and star shapes for $n=2$ and $3$, respectively,  
which is consistent with the implication of $P_s(s,p)$. 
Some of these resonances look like strongly scarred eigenfunctions of billiard system,
showing strong directional emissions matched to the triangular and star patterns.
The $nkR$ values and  patterns for whole resonances will be presented elsewhere
due to lack of space.

 The most clearly localized resonances for $n=2$ and $3$ are shown in Fig.~3.
The patterns look like strongly scarred resonances, but there is no exact
underlying unstable periodic orbit.
Absence of periodic orbits of simple geometry, without bouncing at notch, e.g., triangle and star,
is evident by numerical evaluation of $\delta p =p_i -p_f$ for a closed triangle or star trajectory 
starting from $(s_i,p_i)$ and terminating at $(s_i,~p_f)$. 
Moreover nonexistence of periodic orbits of simple geometry can be understood if one
know that for the clockwise rotating case the distance between origin and the ray segment
always decreases as far as the ray bounces at the curved part of the spiral-shape boundary.
We obtain $  |\delta p| > \sigma$ for arbitrary $s_i$ value, 
where $\sigma=0.075$ for the triangle trajectory and $\sigma=0.136$
for the star trajectory.
Since the localized patterns of resonances are not supported by any unstable periodic orbit,
we call them {\it quasi-scarred resonances}.
The existence of quasi-scarred resonances in dielectric cavities can be understood
from the inherent property of dissipative systems, i.e., uncertainty characteristics.
Another important result from the resonance pattern analysis is that
many resonances are quasi-scarred, e.g., in the present case  more
than a half are 
quasi-scarred, while only a small fraction of eigenfunctions are scarred in billiard systems. 
In practical experiment, this implies that the quasi-scarred lasing
emission can be excited easily due to its dominant existence in resonances.
In fact, the dominant existence of quasi-scarred resonances 
can be regarded as a result of the openness of microcavities.
In open systems, rather local part of phase space would support
resonances(e.g., see Fig. 2(a)) and the resulting individual resonance would 
show a stong localization whose pattern might be determined by the property of 
the openness. This is consistent with results, associated with scarred resonances,
in various open systems~\cite{Ku01,Kim02}
        
\begin{figure}
\begin{center}
\vspace{-0.2 cm}
 \includegraphics[height=3.5cm, width=3.7cm]{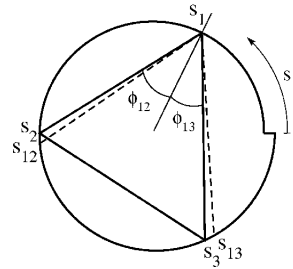}
\vspace{-0.1cm}
\caption{ Schematic diagram for quantifying the degree of uncertainty.
The trajectory satisfying the Snell's law, with an incident angle 
$(\phi_{12}+\phi_{13})/2$, is denoted by dashed lines.}
\vspace{-0.5cm}
\end{center}
\end{figure}

   Now, we consider bouncing positions of the triangle formed in quasi-scarred resonances
which seem to have a definite dependence on their $\mbox{Re}(nkR)$ values.  
We assume that the triangle in quasi-scarred resonances has minimum deviation from the ray
trajectory governed by the Snell's law, and maximum constructive interference under
constraint of high intensity of the electric field at the bouncing positions.
We quantify these by two factors, $\alpha$ and $\beta$ as follows.
Let $s_i$ $(i=1,~2,~3)$ be the bouncing positions of a triangle, from
the angles ($\phi_{ij},~\phi_{ik}$) to the normal line on the
boundary, and we can define $p_{ij}=\sin(\phi_{ij})$,
$p_{ik}=\sin(\phi_{ik})$, and $p_i
=\sin(\frac{\phi_{ij}+\phi_{ik}}{2})$ (here $i,~j,~k$ are
cyclic). Also we get the new positions $s_{ij}, s_{ik}$ as the next
positions of $(s_i,~p_i)$ and $(s_i,~-p_i)$, respectively (see
Fig.~4). Then we define partial uncertainty of the triangle given by
$(s_1,~s_2,~s_3)$ as 
\begin{equation}
\alpha_i = [(p_i-p_{ij})(s_{j}-s_{ij})]^{2}+[(p_i-p_{ik})(s_{k}-s_{ik})]^2. 
\end{equation}
Total uncertainty, therefore, is the sum of these terms, $\alpha =\sum_{i=1}^3 \alpha_i$.
By definition, when the triangle is a periodic orbit, $\alpha$ becomes zero.
To quantify the degree of constructive interference we consider 
$ m_i+\beta_i=l_{i}/(\lambda/2)+\delta \phi/\pi$
for each triangle segment of length $l_i$,
where $m_i$ is an integer and $-0.5< \beta_i < 0.5$,
$\lambda=2\pi/(nk)$, and $\delta \phi$ is the phase shift arisen 
from total internal reflection~\cite{Fo75}.
Total quantity for the degree of constructive interference is then 
$\beta=\sum_{i=1}^3\beta_i^2$ with an additional constraint that the sum $M=\sum_{i=1}^3 m_i$ 
should be even. 

\begin{figure}
\vspace{-0.2cm}
\hspace{-0.5 cm} \includegraphics[height=5cm, width=5.5cm]{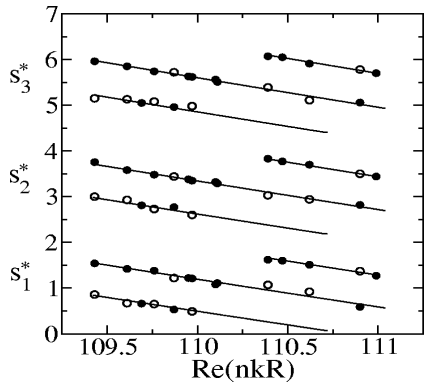} 
\vspace{-0.3cm}
\caption{Variation of the optimized bouncing positions ($s_1^*,~s_2^*,~s_3^*$).
The solid lines denote the present theory with a correction $\mu=0.06$.
The circles represent the bouncing positions of triangular
quasi-scarred resonance patterns of $n=2$ case; three solid circles
with the same $\mbox{Re}(nkR)$ correspond to the main triangular
pattern, and three open circles do to the secondary triangular pattern in a quasi-scarred resonance. }
\vspace{-0.3cm}
\end{figure}

  We first determine triangles with minimum uncertainty $\alpha$ as a function of $s_1$,
and then  apply the condition of minimum $\beta$ to the triangles. From this process
we get the most optimized triangle of ($s_1^*,~s_2^*,~s_3^*$) for a fixed $\mbox{Re}(nkR)$.
The direct application of this method shows systematic deviation from bouncing positions
of resonance patterns. 
This systematic discrepancy results from the fact that
rays inside microcavity have angular distributions and, 
also the boundary has curvatures, which gives rise to a correction of the Snell's law.   
This effect is prominent near the critical angle $\theta_c$,
studied and known as Goos-H\"{a}nchen~\cite{Go47} and
Fresnel Filtering effects~\cite{Re02,Tu02}. We here incorporate these
effects by taking effective segment length $l^*_i=l_i+\mu\lambda$. The
results are shown in Fig.~5. The solid lines are results of the
present theory with $\mu=0.06$ which are in good agreement with the
bouncing positions (denoted by circles) of the quasi-scarred
resonances. Absence of bouncing positions near $s=2.0$ and $s=4.5$ is
consistent with the tentacular structure of the approximate steady
probability distribution $P_s(s,p)$ in Fig.~2 (a).

   In conclusion, we have found that the localized patterns of resonances
in a spiral-shaped dielectric microcavity are constructed by the quasi-scar
phenomenon which comes from inherent properties of the dielectric
microcavity, and that a large fraction of the resonances are
quasi-scarred. The results are contrasted with the case of billiard
systems in which only scar phenomenon exists, and a small fraction of eigenfunctions are scarred.
Even though the system is chaotic, it is possible to extract some information
on resonance patterns from the ray dynamical consideration, more precisely,
from the steady probability distribution $P_s(s,p)$.
Since $P_s(s,p)$ contains long lasting ray dynamical information, its structure should
be related to the high-$Q$ resonances which are likely to appear as lasing
modes. 
From a theoretical viewpoint, just like the semiclassical approach in Hamiltonian systems~\cite{Cr92},
a semiclassical method in dielectric cavities might be useful to understand resonance
positions and degree of scarring or quasi-scarring.
Developing semiclassical theory for dielectric cavities seems to be nontrivial. 
We expect that the results of this Letter will improve physical insight onto resonance
patterns in microcavities.

   This work is supported by Creative Research Initiatives of the Korean Ministry of
Science and Technology. S.-Y. Lee would like to thank S.W. Kim for useful discussion
during the Focus Program of APCTP.


\end{document}